# Evaluating the Reproducibility of Research in Obstetrics and Gynecology

Aaron Bowers, BS[1], Shelby Rauh, MS[1], Drayton Rorah, BS[2], Daniel Tritz, MS[1], Lance Frye[3], Matt Vassar, PhD[1].

1. Oklahoma State University Center for Health Sciences, Tulsa, Oklahoma.
2. Kansas City University of Medicine and Biosciences, Joplin, Missouri.
3. Department of Obstetrics and Gynecology, Oklahoma State University Medical Center, Tulsa, Oklahoma

**Corresponding author:** Mr. Aaron Bowers, Oklahoma State University Center for Health Sciences, 1111 W 17th St., Tulsa, OK 74107, United States.

Email: Aaron.bowers@okstate.edu

**Conflicts of Interest and Source of Funding:** This study was funded through the 2019 Presidential Research Fellowship Mentor – Mentee Program at Oklahoma State University Center for Health Sciences.

**Abstract**

**Objective**: Reproducibility is a core tenet of scientific research. A reproducible study is one where the results can be recreated by different investigators in different circumstances using the same methodology and materials. Unfortunately, reproducibility is not a standard to which the majority of research is currently adherent.

**Methods**: We objectively evaluated 300 trials in the field of Obstetrics and Gynecology for fourteen indicators of reproducibility. These indicators include availability of data, analysis scripts, pre-registration information, study protocols and whether or not the study was available via Open Access. We also assessed the trials for financial conflict of interest statements and source of funding.

**Results**: Of the 300 trials in our sample, 208 contained empirical data that could be assessed for reproducibility. None of the trials in our sample provided a link to their protocols or provided a statement on availability of materials. None were replication studies. Just 10.58% provided a statement regarding their data availability, while only 5.82% provided a statement on preregistration. 25.85% failed to report the presence or absence of conflicts of interest and 54.08% did not state the origin of their funding.

**Conclusion**: Research in the field of Obstetrics and Gynecology is not consistently reproducible and frequently lacks conflict of interest disclosure. Consequences of this could be far-reaching and include increased research waste, widespread acceptance of misleading results and erroneous conclusions guiding clinical decision-making.

**Introduction**

Reproducibility in scientific literature refers to the ability of a third party researcher to replicate results of an original study, using the same materials and methodology.[1] Reproducibility is one of the core tenets of scientific research, important so that authors can repeat experiments done by others and gain additional information to support or contradict the original results.[2]

Unfortunately, due to the race to publication and the pressures to be the first to reveal new findings, reproducibility and study repetition has fallen by the wayside.[3] Inability to reproduce results can increase research waste, lead to widespread acceptance of misleading results and allow for erroneous conclusions to find their way into clinical decision-making.[4]

Oseltamivir, marketed under the trade name of Tamiflu, was first sold by Roche in 1999 and was purchased in the magnitude of tens of millions of doses during the H1N1 scare of 2009.[5] It wasn't until 2014 that the Cochrane Collaboration was able to reproduce and publish a review of all of the data that Roche gathered from clinical trials conducted in the early 2000s.[6,7] Cochrane's review found that the drug only decreased duration of symptoms from 7 days to 6.3 days, with no decrease in the number of hospitalizations or serious complications.[8] In this situation, a pharmaceutical company intentionally discouraged reproducibility, costing governments around the world billions of dollars.[5]

Our study seeks to objectively evaluate the reproducibility of a broad, random sample of literature in the field of Obstetrics and Gynecology. Specifically we look for the indicators of reproducibility outlined by Hardwicke et al, including availability of data, analysis scripts, pre-registration information, study protocols and whether the study was available via Open Access.[9] A study of this type and magnitude has never been conducted on Obstetrics and Gynecology literature, but there is evidence that there is a need for it. In 2018, Holding et al. published research indicating that estrogen receptor-α activation is consistently sustained,[10] contradicting the extensively-cited studies from 2000 and 2003, which indicated that activation was cyclical.[11,12] Holding reports that his team attempted to reproduce and build on the results of the original studies, but obtained an entirely unexpected result, one which could have far-reaching consequences for estrogen receptor treatments down the road.[3] Holding doubts that his team was the first to fail to reproduce the original results, but raises questions about the reproducibility of research and the presence of study repetition in Obstetrics and Gynecology literature.[3] We hope that our findings may serve as a baseline on which reproducibility can be measured and serve as a starting point for discussions on how to improve future research in the field.

**Methods**

This study is observational and utilizes a cross-sectional design. We utilized the methodology of Hardwicke et al.[9] with modifications. We reported our study in accordance with guidelines for meta-epidemiological methodology research[13]. Pertinent materials are available on the Open Science Framework (https://osf.io/n4yh5/).

*Journal and Study Selection*
The National Library of Medicine catalog was used to search journals using the tag; for this project we searched twice with Obstetrics[ST] and Gynecology[ST]. This search was performed on May 29, 2019. Inclusion criteria included "English" and "MEDLINE indexed". Any journals that overlapped between the search terms were only included once. We extracted the ISSN for the final list of journals. If the electronic ISSN was unavailable, the linking ISSN was used. These ISSN were used for our Pubmed search, which was performed on May 31, 2019. Our time-frame for publications was from January 1, 2014 to December 31, 2018. From this PubMed search, we randomly sampled 300 publications to extract data from (https://osf.io/fkj4c/).

*Data Extraction Training*
We held a training meeting to ensure inter-rater reliability for our two investigators (DR and SR). The training included an in-person session to introduce the study protocol and extraction form. Investigators extracted data from 3 example articles independently. The investigators then compared data and reconciled differences. The training session was posted online (https://osf.io/tf7nw/). Finally, investigators extracted data from the first 10 articles on the list and held a brief consensus meeting to answer any final questions and reconcile any additional differences. Investigators then proceeded to extract data on the remaining 294 articles. Following all data extraction, a final consensus meeting was held to resolve any remaining differences. AB was available as a third author to resolve persistent disagreements, if any.

*Data Extraction*

Data extraction was conducted in duplicate and blinded fashion. AB was available as a third author to resolve persistent disagreements, but was not required. Additions were made to the Google Form provided by Hardwicke et al. and the form was pilot-tested. [9] The form was designed so that coders could identify if a given study contained the information necessary to be reproducible. (https://osf.io/3nfa5/). The literature we searched was highly variable, including study designs that were excluded due to lack of empirical data (e.g., editorials, simulations, letters, etc.). Where available, we elected to include the 5-year impact factor and the impact factor for the most recent year. We included the following study designs: Meta-analyses, commentaries with re-analysis, cost-effectiveness studies, clinical trials, case studies, case control studies, data surveys, laboratory studies, multiple study types, cohort studies, case series, secondary analyses, chart reviews, and cross-sectional studies. Funding options included public, private, university, hospital, non-profit, no statement, no funding and mixed.

*Open Access Availability*

The essential components needed to assess a study for its measure of reproducibility are only available within the full text. To determine the ability of the general public to access the full text of each publication, we used the Open Access Button, Google, and PubMed. First, the DOI was used to search for each publication using the Open Access Button to determine if it was freely available to the public. If the Open Access Button returned no results or reported an error, we searched for the publication on Google and PubMed and analyzed the journal website to determine if the publication was available without a paywall.

*Statistical Analysis*

Descriptive statistics will be reported with 95% confidence intervals using Microsoft Excel.

**Results**

Of the 300 articles in our original sample, 2 were excluded due to not being in the English language and 4 were excluded because they could not be accessed. Of the 294 remaining articles, 86 (29.25% [24.10% to 34.40%]) contained no empirical data, leaving 208 studies to be assessed for reproducibility. (Figure 1) The most common study design in our sample was cohort (58/294, 19.73% [15.22% to 24.23%]), followed by cross-sectional (30/294, 10.20% [6.78% to 13.63%]), clinical trial (19/294, 6.46% [3.68% to 9.24%]), laboratory (19/294, 6.46% [3.68% to 9.24%]), case study (16/294, 5.44% [2.88% to 8.01%]) and meta-analysis (15/294, 5.01% [2.61% to 7.59%]). All other study types made up less than 5% each of our sample. (Table 2)

Among studies in our sample, 32 (16.58%) have been cited at least once by a meta-analysis, with 10 (5.18%) being cited in multiple meta-analyses. Median 5-year impact factor of the journals the studies were published in was 2.478 (range .609-5.825).

Studies in our sample had human subjects 59.52% of the time (175/294 [53.97% to 65.08%]), animal subjects 2.04% of the time (6/294 [0.44% to 3.64%]) and the remaining 38.44% (113/294 [32.93% to 43.94%]) had neither human nor animal subjects.

*Indicators of Reproducibility*
Of the 294 articles in our sample, none claimed to be a replication study. In our sample, 72/294 (24.49% [19.62% to 29.36%] were available via the Open Access Button, 2 (0.68% [0% to 1.61%]) were accessible via other means, and 220 (74.83% [69.92% to 79.74%]) were only accessible through a paywall.

Among studies with empiric data, 169 studies (89.42% [85.94% to 92.90%]) failed to provide a statement regarding the availability of their data. 20 (10.58% [7.10% to 14.06%]) provided a statement that data was available, with 13 providing data in a supplement, 1 hosting data on a third-party site and 6 stating that data is available upon request. Only 11 studies (5.82% [3.17% to 8.47%]) provided a statement on pre-registration, 9 of those were accessible. Among those that were accessible, 5 outlined the study's hypothesis, 6 described the methods and 3 mentioned the plan for analysis. 168 (96.55% [94.49% to 98.62%]) studies failed to include a statement on materials availability, 5 (2.87% [0.98% to 4.76%]) reported that materials were available and 1 stated that materials are not available. None of the studies in our sample provided a statement regarding analysis scripts or linked to accessible protocols. (Table 1)

*Conflict of Interest and Funding Statements*

294 articles in our sample were assessed for conflict of interest statements and funding source. 76 (25.85% [20.90% to 30.80%]) contained no conflict of interest statement, 199 (67.69% [62.39% to 72.98%]) contained a statement reporting no financial conflict of interest and the remaining 19 (6.46% [3.68% to 9.24%]) contained a statement reporting one or more conflicts of interest. Among payment sources, mixed funding (38/294, 13.26% [9.43% to 17.10%]) was the most common, followed by public (25/294, 8.50% [5.34% to 11.66%]) and private (21/294, 7.14% [4.23% to 10.06%]). 38 (12.93% [9.13% to 16.72%]) report no funding received and 159 (54.08% [48.44% to 59.72%]) contain no statement regarding the source of funding. The 12 remaining studies received funding from other sources. (Supplemental Table 1)

**Discussion**

Our study identified significant deficiencies in the reproducibility of research in Obstetrics and Gynecology. Unfortunately, our findings are not unique or unexpected. In a similar survey of social science research, Hardwicke et al. found that fewer than 10% of the studies in their sample provided a data availability statement, information about pre-registration, a materials statement, links to protocols or analysis scripts.[9] In addition, they found that less than half of the studies in their sample were available through Open Access and the majority did not include a statement on the source of their funding. All of the above results match with what we found in our survey.

A major area of concern from our sample is the lack of data availability. In bioinformatics and biostatistics, Ioannidis et al. attempted to reproduce 16 studies and found that 2 were reproducible, 6 were partially reproducible and 10 were not reproducible, with the main reason for failure being lack of data availability.[14] Other studies have found that many journals do not have a policy regarding data availability.[15,16] While journals that do require data availability have increased data availability, sub-optimal adherence continues to be an issue.[17] We found that fewer than 11% of studies in our sample provided a data availability statement. While it would be ideal for all studies to provide their data in a supplement or to be hosted on an accessible third-party site, authors should take it upon themselves to at least make data available upon reasonable request. Inability to access raw data can lead to inaccuracies and misleading results in the medical literature.[18]

Compounding the lack of data availability is the lack of data accessibility. Our sample found that only one in four articles was available through Open Access or with our academic privilege. If investigators are unable to access relevant research to their topic, it is unreasonable to expect that research to be reproduced. Indeed we found no articles that claimed to replications of previous studies. A survey by Baker et al. found that while a majority of researchers had failed to reproduce a study, few researchers attempted to publish their findings.[19] The same survey reported that the most common reasons for poor reproducibility are selective reporting, pressure to publish, poor analysis, low statistical power and the research not being well-replicated in the original lab.

In our data set, no studies provided links to protocols or analysis scripts and only 6 contained a materials statement. Protocols, materials statements and analysis scripts are three of the most critical elements of reproducibility in a study. Protocols provide a painstakingly thorough guide to how each step of the study was performed, containing details that may not be relevant to most of the individuals reading the methods section.[20] Analysis scripts provide explicit details on how the analysis was conducted, while materials statements allow future researchers to replicate the experiment using the same tools, questionnaires, forms, etc.[21] Simply requiring that protocols, analysis scripts and materials statements are provided in supplement or on a third-party website is an easy way for journal editors to improve the reproducibility of the studies they publish.

Only 11 studies in our sample provided a statement on pre-registration. Nosek et al. underscores the importance of pre-registration by distinguishing between hypothesis-generating research and hypothesis-confirming research, noting that the two should not occur simultaneously.[22] Crucially,

Nosek states that presenting post hoc explanations as predictions decreases reproducibility and falsely inflates the authors conviction in the strength of their evidence. This is to say nothing of p-hacking or selective outcome reporting, both of which can be easily identified in pre-registered studies.[23,24] Over the last several decades, the Food and Drug Administration (FDA) and the World Health Organization (WHO) have passed legislation to increase transparency and reproducibility in clinical trials by creating the Clinicaltrials.gov and the International Clinical Trials Registry Platform. [25,26] Since 2007, the United States has mandated the pre-registration of clinical trials through the FDA Amendments Act.[27] While not a total success, clinical trial registration has improved significantly.[28,29] While government policy could be an effective way to encourage the pre-registration of all empiric research, at this point it is up to authors to pre-register their studies and journal editors to implement policies that every study be pre-registered.

Among the 294 studies in our sample, only 218 contained a statement regarding financial conflict of interest. Once again, this is not a unique issue in research in Obstetrics and Gynecology. Cross-sectional studies in several fields have shown significant deficits in conflict of interest reporting, where 100% disclosure is the only acceptable outcome.[30–33] Thompson et al. found that among abstract authors in Obstetrics and Gynecology, 62% had undisclosed financial conflicts of interest, with the majority of those (68%) being relevant to the topic of the abstract.[34] Financial conflicts of interest are a rampant problem in scientific literature, and it is clear that self-reporting is not sufficient as a method to fix the problem.[35–37]

To help solve the problems of poorly reproducible research, The Open Science Framework (OSF) was launched by the Center for Open Science.[38] One of the ways they set out to accomplish this was to host research pre-registration free of charge. Additionally, in 2015, the OSF published the Transparency and Openness Promotion (TOP) guidelines which outlined 8 modular standards across 3 tiers to improve transparency, cooperation and reproducibility.[39] Separately, McIntosh et al. developed Repeat, a checklist for authors that emphasizes 5 areas of reproducibility, including many of the same indicators that our study assessed for.[40] Reporting guidelines are another tool that some journals require to ensure that data is reported completely and accurately.[41–43] More regular use of any of the above tools would almost certainly lead to increased reproducibility.

Our study contains several important limitations. First is our sample size compared to the bulk of research in Obstetrics and Gynecology. We felt that 300 studies would be a suitably robust sample to give us a good snapshot of the reproducibility of literature in the field, but we would be naive to believe that our sample was so large that it could be definitively labeled as "representative." Second, we did not individually interrogate individual authors for data availability, analysis scripts or protocols. Research has already been done on the success rates of author inquiry and we felt it wasn't necessary to include in our analysis.[44,45]

**Conclusion**

Research in the field of Obstetrics and Gynecology rates poorly on indicators of reproducibility. Additionally, there are an unacceptable number of studies published without financial conflict of interest disclosures. We call upon authors to take steps to increase reproducibility including using reporting guidelines, the Repeat checklist and by pre-registering their studies. Additionally we recommend professional societies and journal editors to adopt more stringent policies regarding data and protocol availability as well as conflicts of interest statements.

**References**

1. Plesser HE. Reproducibility vs. Replicability: A Brief History of a Confused Terminology. *Frontiers in Neuroinformatics*. 2018;11. doi:10.3389/fninf.2017.00076

2. Joshi M, Bhardwaj P. Impact of data transparency: Scientific publications. *Perspect Clin Res*. 2018;9(1):31-36.

3. Holding AN. Novelty in science should not come at the cost of reproducibility. *FEBS J*. 2019;103:843.

4. Ioannidis JPA, Greenland S, Hlatky MA, et al. Increasing value and reducing waste in research design, conduct, and analysis. *Lancet*. 2014;383(9912):166-175.

5.  Jack A. Tamiflu: "a nice little earner." *BMJ*. 2014;348(apr09 2):g2524-g2524. doi:10.1136/bmj.g2524

6.  Payne D. Tamiflu: the battle for secret drug data. *BMJ*. 2012;345:e7303.

7.  Cohen D. Complications: tracking down the data on oseltamivir. *BMJ*. 2009;339:b5387.

8.  Jefferson T, Jones MA, Doshi P. Neuraminidase inhibitors for preventing and treating influenza in adults and children. *Cochrane Database Syst Rev*. 2014. https://www.cochranelibrary.com/cdsr/doi/10.1002/14651858.CD008965.pub4/full.

9.  Hardwicke TE, Wallach JD, Kidwell M, Ioannidis J. An empirical assessment of transparency and reproducibility-related research practices in the social sciences (2014-2017). April 2019. doi:10.31222/osf.io/6uhg5

10. Holding AN, Cullen AE, Markowetz F. Genome-wide Estrogen Receptor-α activation is sustained, not cyclical. *Elife*. 2018;7. doi:10.7554/eLife.40854

11. Shang Y, Hu X, DiRenzo J, Lazar MA, Brown M. Cofactor Dynamics and Sufficiency in Estrogen Receptor–Regulated Transcription. *Cell*. 2000;103(6):843-852.

12. Métivier R, Penot G, Hübner MR, et al. Estrogen receptor-alpha directs ordered, cyclical, and combinatorial recruitment of cofactors on a natural target promoter. *Cell*. 2003;115(6):751-763.

13. Murad MH, Wang Z. Guidelines for reporting meta-epidemiological methodology research. *Evid Based Med*. 2017;22(4):139-142.

14. Ioannidis JPA, Allison DB, Ball CA, et al. Repeatability of published microarray gene expression analyses. *Nat Genet*. 2009;41(2):149-155.

15. Alsheikh-Ali AA, Qureshi W, Al-Mallah MH, Ioannidis JPA. Public availability of published research data in high-impact journals. *PLoS One*. 2011;6(9):e24357.

16. Hardwicke TE, Mathur MB, MacDonald K, et al. Data availability, reusability, and analytic reproducibility: evaluating the impact of a mandatory open data policy at the journal Cognition. *R Soc Open Sci*. 2018;5(8):180448.

17. Federer LM, Belter CW, Joubert DJ, et al. Data sharing in PLOS ONE: An analysis of Data Availability Statements. *PLoS One*. 2018;13(5):e0194768.

18. Steinbrook R, Kassirer JP. Data availability for industry sponsored trials: what should medical journals require? *BMJ*. 2010;341:c5391.

19. Baker M. 1,500 scientists lift the lid on reproducibility. *Nature*. 2016;533(7604):452-454.

20. Jirge PR. Preparing and Publishing a Scientific Manuscript. *J Hum Reprod Sci*.

2017;10(1):3-9.

21. Piccolo SR, Frampton MB. Tools and techniques for computational reproducibility. *Gigascience*. 2016;5(1):30.

22. Nosek BA, Ebersole CR, DeHaven AC, Mellor DT. The preregistration revolution. *Proc Natl Acad Sci U S A*. 2018;115(11):2600-2606.

23. Howard B, Scott JT, Blubaugh M, Roepke B, Scheckel C, Vassar M. Systematic review: Outcome reporting bias is a problem in high impact factor neurology journals. *PLoS One*. 2017;12(7):e0180986.

24. Dodson TB. The Problem With P-Hacking. *Journal of Oral and Maxillofacial Surgery*. 2019;77(3):459-460. doi:10.1016/j.joms.2018.12.034

25. Food, Administration D, Others. Food and Drug Administration modernization act (FDAMA) of 1997. 2013.

26. Organization WH, Others. WHO statement on public disclosure of clinical trial results. 2015.

27. United States. Congress. House. Committee on Energy and Commerce. *Food and Drug Administration Amendments Act of 2007: Report Together with Additional Views (to Accompany H.R. 2900) (including Cost Estimate of the Congressional Budget Office)*. U.S. Government Printing Office; 2007.

28. Zou CX, Becker JE, Phillips AT, et al. Registration, results reporting, and publication bias of clinical trials supporting FDA approval of neuropsychiatric drugs before and after FDAAA: a retrospective cohort study. *Trials*. 2018;19(1). doi:10.1186/s13063-018-2957-0

29. Phillips AT, Desai NR, Krumholz HM, Zou CX, Miller JE, Ross JS. Association of the FDA Amendment Act with trial registration, publication, and outcome reporting. *Trials*. 2017;18(1):333.

30. Rasmussen K, Schroll J, Gøtzsche PC, Lundh A. Under-reporting of conflicts of interest among trialists: a cross-sectional study. *J R Soc Med*. 2015;108(3):101-107.

31. Dunn AG, Coiera E, Mandl KD, Bourgeois FT. Conflict of interest disclosure in biomedical research: A review of current practices, biases, and the role of public registries in improving transparency. *Res Integr Peer Rev*. 2016;1. doi:10.1186/s41073-016-0006-7

32. Kesselheim AS, Wang B, Studdert DM, Avorn J. Conflict of interest reporting by authors involved in promotion of off-label drug use: an analysis of journal disclosures. *PLoS Med*. 2012;9(8):e1001280.

33. Roseman M, Turner EH, Lexchin J, Coyne JC, Bero LA, Thombs BD. Reporting of conflicts of interest from drug trials in Cochrane reviews: cross sectional study. *BMJ*.

2012;345:e5155.

34. Thompson JC, Volpe KA, Bridgewater LK, et al. Sunshine Act: shedding light on inaccurate disclosures at a gynecologic annual meeting. *Am J Obstet Gynecol*. 2016;215(5):661.e1-e661.e7.

35. Vassar M, Bibens M, Wayant C. Transparency of industry payments needed in clinical practice guidelines. *BMJ Evid Based Med*. 2019;24(1):8-9.

36. Carlisle A, Bowers A, Wayant C, Meyer C, Vassar M. Financial Conflicts of Interest Among Authors of Urology Clinical Practice Guidelines. *Eur Urol*. 2018;74(3):348-354.

37. Checketts JX, Sims MT, Vassar M. Evaluating Industry Payments Among Dermatology Clinical Practice Guidelines Authors. *JAMA Dermatol*. 2017;153(12):1229-1235.

38. OSF. https://osf.io/. Accessed July 12, 2019.

39. Nosek BA, Alter G, Banks GC, et al. Transparency and Openness Promotion (TOP) Guidelines. October 2016. doi:10.1126/science.aab2374

40. McIntosh LD, Juehne A, Vitale CRH, et al. Repeat: a framework to assess empirical reproducibility in biomedical research. *BMC Med Res Methodol*. 2017;17(1):143.

41. Checketts JX, Cook C, Imani S, Duckett L, Vassar M. An Evaluation of Reporting Guidelines and Clinical Trial Registry Requirements Among Plastic Surgery Journals. *Ann Plast Surg*. May 2018. doi:10.1097/SAP.0000000000001476

42. Jorski A, Scott J, Heavener T, Vassar M. Reporting guideline and clinical trial registration requirements in gastroenterology and hepatology journals. *Int J Evid Based Healthc*. 2018;16(2):119-127.

43. Sims MT, Bowers AM, Fernan JM, Dormire KD, Herrington JM, Vassar M. Trial registration and adherence to reporting guidelines in cardiovascular journals. *Heart*. 2018;104(9):753-759.

44. Vines TH, Albert AYK, Andrew RL, et al. The availability of research data declines rapidly with article age. *Curr Biol*. 2014;24(1):94-97.

45. Hardwicke TE, Ioannidis JPA. Populating the Data Ark: An attempt to retrieve, preserve, and liberate data from the most highly-cited psychology and psychiatry articles. *PLoS One*. 2018;13(8):e0201856.

**Figure 1. Prisma Diagram for included and excluded studies**

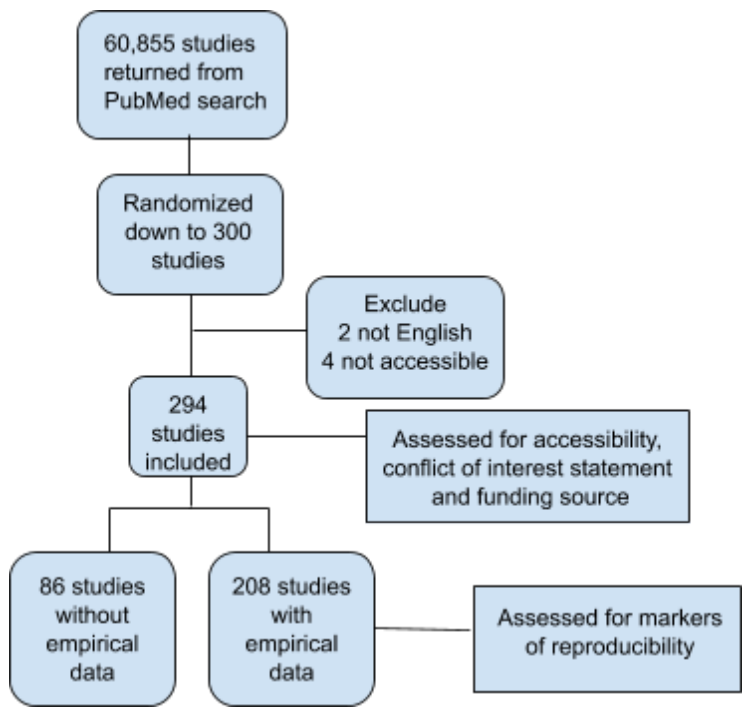

**Table 1:** Reproducibility related characteristics. Variable numbers (N) are dependent upon study design. Full detailed protocol pertaining to our measured variables is available online (https://osf.io/x24n3/)

| *Indicators of Reproducibility Included in Present Study* | *Significance of measure variable for transparency and reproducibility.* |
|---|---|
| **Publications** | |

| | Publication accessibility (Is the publication open access to the general public or accessible through a paywall?) | The general public's ability to access scientific research may increase transparency of results and improve the ability for others to critically assess studies, potentially resulting in more replication studies |
| --- | --- | --- |
| All (N=300) | | |
| **Funding** | | |
| Included studies (N=294) | Funding statement (Does the publication state their funding sources?) | Explicitly providing source of funding may help mitigate bias and potential conflicts of interest |
| **Conflict of Interest** | | |
| Included studies (N=294) | Conflict of interest statement (Does the publication state whether or not the authors had a conflict of interest?) | Explicitly providing conflicts of interest may allow for full disclosure of factors that may promote bias in the study design or outcomes |
| **Publication Citations** | | |
| Empirical studies† (N=208) | Citations by a systematic review/meta-analysis (Has the publication been cited by any type of data synthesis publication, and if so, was it explicitly excluded?) | Systematic reviews and meta-analyses evaluate and compare existing literature to assess for patterns, strengths, and weaknesses of studies regarding a particular field or topic |
| **Analysis Scripts** | | |
| Empirical studies‡ (N=189) | Availability statement (Does the publication state whether or not the analysis scripts are available?) | Access to the analysis script helps promote credibility by ensuring raw data can be analyzed in the same way to get the same |
| | Method of availability (Ex: Are the analysis scripts available upon request or in a supplement?) | |
| | Accessibility (Can you view, download, or otherwise access the analysis scripts?) | |
| **Materials** | | |
| Empirical studies¶ (N=174) | Availability statement (Does the publication state whether or not the materials are available?) | Access to the study materials helps promote credibility by ensuring the study can be conducted in the same way, using the same materials |
| | Method of availability (Ex: Are the materials available upon request or in a supplement?) | |
| | Accessibility (Can you view, download, or otherwise access the materials?) | |
| **Pre-registration** | | |
| Empirical studies‡ (N=189) | Availability statement (Does the publication state whether or not it was pre-registered?) | Pre-registering studies may decrease potential bias and increase the overall credibility of a study |
| | Method of availability (Where was the publication pre-registered?) | |
| | Accessibility (Can you view or otherwise access the registration?) | |
| | Components (What components of the publication were pre-registered?) | |
| **Protocols** | | |
| Empirical studies‡ (N=189) | Availability statement (Does the publication state whether or not a protocol is available?) | Access to detailed protocols allows for replication of studies with the same methodology, increasing both the credibility of a study and chance of replication |
| | Components (What components are available in the protocol?) | |
| **Raw Data** | | |
| Empirical studies‡ (N=189) | Availability statement (Does the publication state whether or not the raw data are available?) | Access to raw data can help improve the replication ability of a study and minimize potential for bias and issues of validity and reliability |
| | Method of availability (Ex: Are the raw data available upon request or in a supplement?) | |
| | Accessibility (Can you view, download, or otherwise access the raw data?) | |
| | Components (Are all the necessary raw data to reproduce the study available?) | |
| | Clarity (Are the raw data documented clearly?) | |
| † 'Empirical studies' are publications that include empirical data such as: clinical trial, cohort, case series, case reports, case-control, secondary analysis, chart review, commentaries (with data analysis), laboratory, and cross-sectional designs. ‡ Empirical studies determined to be case reports or case series were excluded in regards to reproducibility related questions (materials, data, protocol, and registration were excluded) as recommended by Wallach et al. (cite) ¶ Empirical studies determined to be either case reports, case series, commentaries with analysis, meta-analysis or systematic review were excluded as they did not provide materials to fit the category. | | |

**Table 2.** Characteristics of Included Publications

| Table 2: Characteristics of Included Publications |||| 
| Characteristics || Variables ||
| | | N (%) | 95% CI |
| **Funding** **N=294** | University | 6 (2.04) | 0.44-3.64% |
| | Hospital | 2 (0.68) | 0-1.61% |
| | Public | 25 (8.50) | 5.35-11.66% |
| | Private/Industry | 21 (7.14) | 4.23-10.06% |
| | Non-Profit | 4 (1.36) | 0.05-2.67% |
| | Mixed | 39 (13.27) | 13.54-14.04% |
| | No Statement Listed | 159 (54.08) | 48.44-59.72% |
| | No Funding Received | 38 (12.93) | 9.13-16.72% |
| | | | |
| **Type of Study** **N=294** | No Empirical Data | 86 (29.25) | 24.10-34.40% |
| | Meta-Analysis | 15 (5.10) | 2.61-7.59% |
| | Cost-Effectiveness | 6 (2.04) | 0.44-3.64% |
| | Clinical Trial | 19 (6.46) | 3.68-9.24% |
| | Case Study | 16 (5.44) | 2.88-8.01% |
| | Case Series | 3 (1.02) | 0-2.16% |
| | Cohort | 58 (19.73) | 15.22-24.23% |
| | Chart Review | 10 (3.40) | 1.35-5.45 |
| | Case Control | 13 (4.42) | 2.10-6.75% |
| | Survey | 8 (2.72) | 0.88-4.56% |
| | Cross-Sectional | 30 (10.20) | 6.78-13.63% |
| | Secondary Analysis | 10 (3.40) | 1.35-5.45) |
| | Laboratory | 19 (6.46) | 3.68-9.24% |
| | Multiple Study Types | 1 (0.34) | 0-1.00% |
| | | | |
| **5 Year Impact Factor** **N=268** | Median | 2.478 | - |
| | 1st Quartile | 1.741 | - |
| | 3rd Quartile | 3.284 | - |
| | Interquartile Range | 1.741-3.284 | - |